\documentclass[10pt,aps,prb,twocolumn,preprintnumbers,amsmath,amssymb,floatfix,showpacs,superscriptaddress]{revtex4-1}
\usepackage[latin1]{inputenc}
\usepackage[T1]{fontenc}
\usepackage[english]{babel}
\usepackage[dvips]{graphicx}
\usepackage{amsmath}
\usepackage{times}
\usepackage{color}

\renewcommand{\AA}{\text{\r{A}}}

\newcommand\Multp{\cdot}
\newcommand\Vek[1]{\vec{#1}}

\begin{document}

\title
{
\boldmath
Design of $n$- and $p$-type oxide thermoelectrics in LaNiO$_3$/SrTiO$_3(001)$ superlattices\\exploiting interface polarity
}

\author{Benjamin Geisler}
\email{Benjamin.Geisler@uni-due.de}
\affiliation{Fakult\"at f\"ur Physik, Universit\"at Duisburg-Essen and Center of Nanointegration (CENIDE), Campus Duisburg, Lotharstr.~1, 47048 Duisburg, Germany}
\author{Ariadna Blanca-Romero}
\affiliation{Department of Chemistry, Imperial College London, London, SW7 2AZ, United Kingdom}
\author{Rossitza Pentcheva}
\email{Rossitza.Pentcheva@uni-due.de}
\affiliation{Fakult\"at f\"ur Physik, Universit\"at Duisburg-Essen and Center of Nanointegration (CENIDE), Campus Duisburg, Lotharstr.~1, 47048 Duisburg, Germany}

\date{\today}

\begin{abstract}
We investigate the structural, electronic, transport, and thermoelectric properties
of LaNiO$_3$/SrTiO$_3(001)$ superlattices containing
either exclusively $n$- or $p$-type interfaces or coupled interfaces of opposite polarity
by using density functional theory calculations with an on-site Coulomb repulsion term.
The results show that significant octahedral tilts are induced in the SrTiO$_3$ part of the superlattice.
Moreover, the La-Sr distances and Ni-O out-of-plane bond lengths at the interfaces
exhibit a distinct variation by about $7\,\%$ with the sign of the electrostatic doping.
In contrast to the much studied LaAlO$_3$/SrTiO$_3$ system, the charge mismatch at the interfaces
is exclusively accommodated within the LaNiO$_3$ layers, whereas the interface polarity
leads to a band offset and to the formation of an electric field within the coupled superlattice.
Features of the electronic structure indicate an orbital-selective quantization of quantum well states.
The potential- and confinement-induced multiband splitting results in complex cylindrical Fermi surfaces
with a tendency towards nesting that depends on the interface polarity.
The analysis of the thermoelectric response reveals a particularly large positive Seebeck coefficient ($135~\mu$V/K)
and a high figure of merit ($0.35$)
for room-temperature cross-plane transport in the $p$-type superlattice that is attributed to the participation of the SrTiO$_3$ valence band.
Superlattices with either $n$- or $p$-type interfaces show cross-plane Seebeck coefficients of opposite sign
and thus emerge as a platform to construct an oxide-based thermoelectric generator
with structurally and electronically compatible $n$- and $p$-type oxide thermoelectrics.
\end{abstract}

\pacs{73.20.-r, 68.65.Cd, 73.40.-c, 73.50.Lw}

\maketitle

\section{Introduction}

Advances in layer-by-layer growth techniques on the atomic scale have made it possible
to design artificial transition metal oxide heterostructures
with specific interfaces that display exotic characteristics notably different from their bulk components.~\cite{Hwang:12, Mannhart:10, Chakhalian:12}
In particular, perovskites in the $AB$O$_3$ structure
like the Pauli-paramagnetic correlated metal~\cite{LNO-STO-FermiSurfaces:09, Ouellette:10, Sreedhar:92, Zhou:03} LaNiO$_3$ (LNO)
are currently of considerable interest in this context.
For instance, epitaxial superlattices (SLs) of LNO and the band insulator LaAlO$_3$ (LAO)
exhibit a confinement-driven metal-to-insulator transition~\cite{ABR:11, Freeland:11}
as well as magnetic order,~\cite{Frano:13, Boris:11, Puggioni:12} that is also sensitive to strain.~\cite{ABR:11}
Electrical transport measurements showed an enhanced sheet conductivity
for LNO/SrTiO$_3(001)$ SLs on (LaAlO$_3$)$_{0.3}$(Sr$_2$AlTaO$_6$)$_{0.7}$ (LSAT),~\cite{Son:10, Hwang:13}
whereas standing-wave excited photoemission experiments and \textit{ab initio} calculations
reported a reduction of the electronic density of states (DOS) at the Fermi energy on LSAT and SrTiO$_3$ (STO).~\cite{Kaiser:11, Han:14, Eiteneer:15}

The LNO/STO$(001)$ SLs bear two major differences to commonly studied oxide interfaces.
Unlike the nonpolar LNO/LAO$(001)$ SLs,
in LNO/STO$(001)$ SLs there is a charge discontinuity at the interface depending on the layer stacking
that can lead to either $n$-type [(LaO)$^{+}$/(TiO$_2$)$^{0}$] or $p$-type [(NiO$_2$)$^{-}$/(SrO)$^{0}$] $\delta$-doping.
However, in contrast to the polar LAO/STO$(001)$ system that comprises two band insulators,
here we are dealing with a combination of a correlated metal and a band insulator.
Our results show that the formal charge mismatch at the LNO/STO$(001)$ interfaces is solely compensated within LNO,
which distinguishes our SLs fundamentally from the LAO/STO$(001)$ system
where the electrostatic doping occurs within STO.
Thus, LNO/STO emerges as an interesting extension to the commonly studied systems
that widens our perspective on the behavior that can be realized in oxide SLs.

In this work, we analyze the influence of the interface polarity in LNO/STO$(001)$ SLs
on the lattice and electronic structure
and the implications for the electronic transport and thermoelectric properties.
We consider SLs with either $n$-, \mbox{$p$-,} or coupled $n$- and $p$-type interfaces.
In the latter case, the SL in total is undoped, but our results indicate the formation of an internal electric field.
An interesting structural feature are the significant octahedral tilts induced in the STO region,
which are uncommon for the bulk compound.
Moreover, our results indicate that the distinct La-Sr distances and Ni-O out-of-plane bond lengths at the interfaces
observed in transmission electron microscopy studies~\cite{Hwang:13, ZhangKeimer:14}
are a result of the electrostatic doping and can be used as a fingerprint of the interface polarity.

\begin{figure*}[tb]
	\centering
	\includegraphics{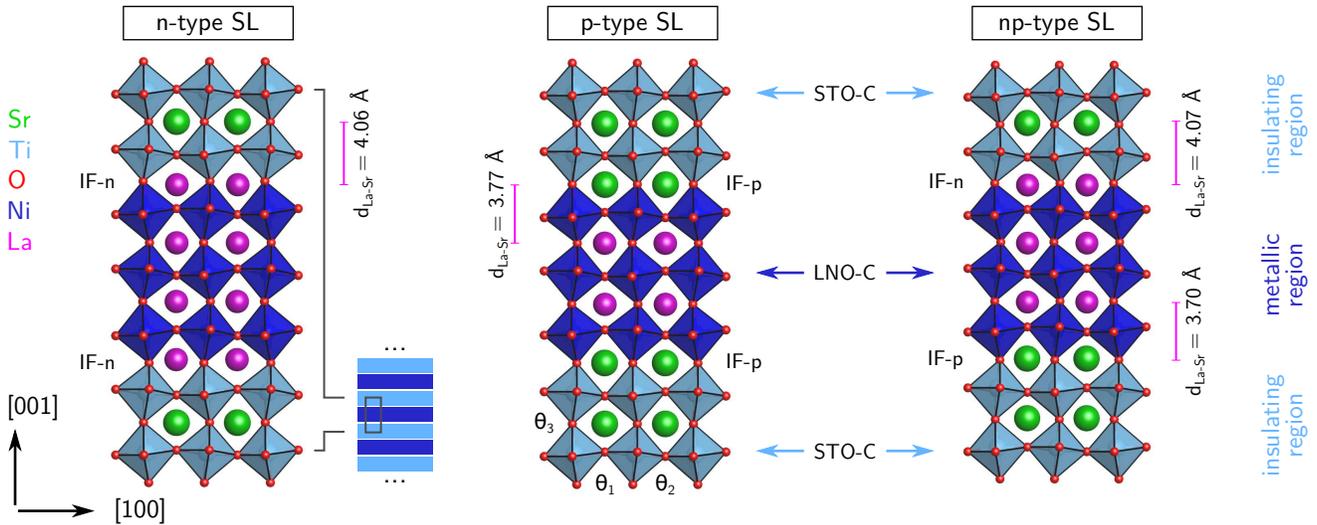}
	\caption{(Color online) Side view of the different optimized SL geometries considered here: an $n$-type SL with two $n$-type interfaces (IF-$n$), a $p$-type SL with two $p$-type interfaces (IF-$p$), and an $np$-type SL where IF-$n$ is coupled with IF-$p$. Purple, green, and small red spheres denote La, Sr, and O ions, respectively, while light blue and dark blue octahedra are centered around Ti and Ni ions, respectively. STO-C and LNO-C mark the central layers in the STO and LNO region, respectively. The directions refer to the \mbox{(pseudo-)}cubic perovskite structure.}
	\label{fig:AtomicStructure}
\end{figure*}

Oxides receive increasing attention for thermoelectric applications due to their chemical and thermal stability and environmental friendliness.~\cite{HeLiuFunahashi:11, HebertMaignan:10}
Considerable research aims at finding oxide thermoelectrics with improved performance, mostly among bulk materials~\cite{Ong:10, Klie:12}
by doping~\cite{Xing:16,Garrity:16,LamontagneGrunerPentcheva:16} or strain.~\cite{Gruner:15}
On the other hand, nanostructuring is expected to significantly improve the thermoelectric properties of a material.~\cite{HicksDresselhaus:93}
The early proposal of Hicks and Dresselhaus has been demonstrated also for oxide SLs.~\cite{Ohta:07, Stemmer:10}
The implications of reduced dimensions have been addressed from first principles only in a few cases,
e.g., for LAO/STO and $\delta$-doped STO SLs~\cite{Pallecchi:15, Filippetti:12, Delugas:13}.
Beyond the effect of confinement, we follow here a different strategy and explore whether $n$- and $p$-type oxide thermoelectrics
can be designed in oxide SLs with selective polarity of their interfaces.  
Hence, we calculate and analyze the in- and cross-plane electronic transport in the aforementioned LNO/STO SLs
by using Boltzmann theory in the constant relaxation time approximation
and provide Seebeck coefficients and estimates for the figure of merit.
We show that the targeted interface design allows to achieve either $n$- or $p$-type thermoelectric response,
which opens a route for constructing an oxide-based thermoelectric generator with compatible $n$- and $p$-type materials.

\section{Methodology}

We have performed first-principles calculations in the framework of spin-polarized density functional theory \cite{KoSh65} (DFT)
employing all-electron and pseudopotential methods.
Exchange and correlations have been described by the generalized gradient approximation (GGA) as parametrized by Perdew, Burke, and Ernzerhof.~\cite{PeBu96}
As in our previous work~\cite{ABR:11} and related literature,~\cite{KimHan:15, Doennig:15, May:10}
we account for correlation effects by using the DFT$+U$ formalism,~\cite{Anisimov:93} setting $U=4$ and $J=0.7$~eV for Ni~$3d$ and Ti~$3d$ (all-electron and pseudopotential) and $U=7$ and $J=0$~eV for La~$4f$ (all-electron).
We checked the robustness of our results by testing different values for $U$ on the Ni and Ti sites ranging from $3$ to $5$~eV
and found only minor quantitative differences in the relevant part of the electronic structure.

In order to take octahedral tilts fully into account,
we model the LNO/STO SLs by using $\sqrt{2}a \times \sqrt{2}a \times c$ supercells,
rotated by $45^\circ$ about the $[001]$ axis with respect to the \mbox{(pseudo-)}cubic perovskite unit cell,
that contain $60$ atoms in total.
We consider the case of SLs grown on an STO$(001)$ substrate
by setting the in-plane lattice parameter to the experimental lattice constant of STO, $a = 3.905~\AA$,
and we use $c = 3 \times 3.905~\AA + 3 \times 3.81~\AA$
following x-ray diffraction results for epitaxially strained LNO films on STO$(001)$.~\cite{May:10}

All-electron calculations have been performed by using the full-potential linearized augmented plane wave plus local orbital (LAPW) technique as implemented in the Wien2k code. \cite{BlSc01}
The muffin tin radii have been set to $0.95~\AA$ (Ni and Ti), $1.22~\AA$ (La and Sr), and $0.85~\AA$ (O).
Wave functions have been expanded inside the muffin tins in spherical harmonics up to $l_{\text{max}}^{\text{wf}}=10$.
Non-spherical contributions to the electron density and potential have been considered up to $l_{\text{max}}^{\text{pot}}=4$.
The plane wave cutoff energy in the interstitial region has been set to $E_{\text{max}}^{\text{wf}}=19$~Ry for the wave functions
and $E_{\text{max}}^{\text{pot}}=144$~Ry for the potential.
A $8 \times 8 \times 1$ Monkhorst-Pack $\Vek{k}$-point grid~\cite{MoPa76} together with the tetrahedron method~\cite{Bloechl:94} has been used to sample the Brillouin zone.
The atomic positions have been optimized until the maximum component of the residual forces on the ions was less than $5$~mRy/a.u.

Converged transport properties require electronic structure data on a very dense $\Vek{k}$-point grid.
Therefore, further investigations have been carried out by using the Quantum Espresso code~\cite{PWSCF} which employs plane waves as basis functions.
Wave functions and density have been expanded into plane waves up to cutoff energies of $25$ and $250$~Ry, respectively.
Ultrasoft pseudopotentials (USPPs) have been used, \cite{Vanderbilt:1990}
treating the
La $5s$, $5p$, $5d$, $6s$, $6p$,
Sr $4s$, $4p$, $4d$, $5s$, $5p$,
Ni $3d$, $4s$,
Ti $3s$, $3p$, $3d$, $4s$, $4p$,
and O $2s$, $2p$
atomic subshells as valence states.
For La, Sr, and Ni a non-linear core correction~\cite{LoFr82} has been included.
A Methfessel-Paxton smearing~\cite{MePa89} of $10$~mRy has been used during the Brillouin zone sampling.
We have found that the USPP results agree very well with the all-electron LAPW results.

The electronic transport properties of the SLs have been calculated from the DFT electronic structure
by using Boltzmann transport theory in the constant relaxation time approximation.
The BoltzTraP code~\cite{BoltzTraP:06} provides the energy- and spin-resolved transmission ${\cal T}_\sigma(E)$.
Several test calculations have shown that
a very dense $64 \times 64 \times 8$ $\Vek{k}$-point grid is required in order to obtain converged transmission curves.
From these we have calculated the thermoelectric quantities
as described in the Appendix and used in previous studies.~\cite{Geisler-Heusler:15, Gruner:15, GeislerPopescu:14, ComtesseGeisler:14, LamontagneGrunerPentcheva:16}

\section{Structural relaxations}

By varying the stacking sequence at the interface we have generated three types of SLs:
an electron-doped one with $n$-type (LaO)$^{+}$/(TiO$_2$)$^{0}$ interfaces (IF-$n$),
a hole-doped one with (NiO$_2$)$^{-}$/(SrO)$^{0}$ interfaces (IF-$p$),
and one with coupled interfaces of opposite polarity.
The resulting stoichiometries are
(LNO)$_{3.5}$/(STO)$_{2.5}$ ($n$-type SL),
(LNO)$_{2.5}$/(STO)$_{3.5}$ ($p$-type SL), and
(LNO)$_{3}$/(STO)$_{3}$ ($np$-type SL).
While the first two SLs are symmetric and doped by $\pm 1/2 e$ per interface and $1 \times 1$ unit area,
the third one is undoped but contains a built-in electric field.
The optimized SL geometries are shown in Fig.~\ref{fig:AtomicStructure}.

The LNO region exhibits an antiferrodistortive $a^{-}a^{-}c^{-}$ octahedral tilting pattern for all considered SLs.
It is a remarkable feature that this tilting pattern extends in the STO region, despite the fact that STO is cubic in the bulk.
To prove the robustness of this feature we have performed additional calculations where the TiO$_6$ octahedral tilts
were initially suppressed and released this constraint only in the final optimization cycles.
The resulting structure with much weaker tilts (not shown here) is $180$~meV (LAPW) per supercell less stable
than the ground state shown in Fig.~\ref{fig:AtomicStructure}.
The octahedral tilts within STO are related to the octahedral connectivity across the interface
and have been observed recently for LNO/STO SLs on LSAT (lattice constant $3.87~\AA$)~\cite{Hwang:13},
but also for LAO/STO SLs~\cite{Doennig:15} and GdTiO$_3$/STO SLs.~\cite{MoetakefStemmer:12}

\begin{figure}[tb]
	\centering
	\includegraphics[]{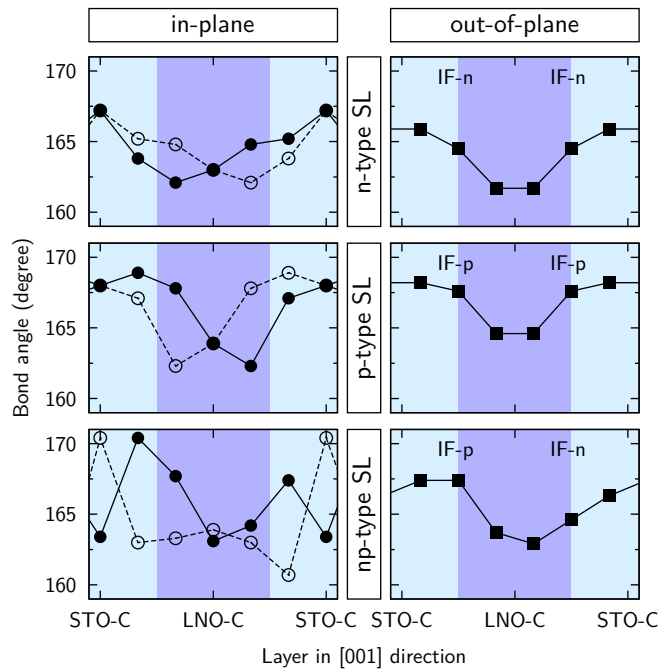}
	\caption{(Color online) Layer-resolved in-plane and out-of-plane $B$-O-$B$ bond angles for the considered SLs. In-plane, the bond angles take on two distinct values, $\theta_1$ and $\theta_2$ (cf.~Fig.~\ref{fig:AtomicStructure}), that alternate along the $[001]$ and $[010]$ directions and are depicted by open and filled circles, respectively. Out-of-plane refers to the $[001]$ direction ($\theta_3$). The inversion symmetry of the $n$- and the $p$-type SL can clearly be seen. The bulk values are $180^\circ$ (STO) and $165^{\circ}/159^\circ$ (in-plane/out-of-plane in strained bulk LNO~\cite{May:10}).}
	\label{fig:BondAngles}
\end{figure}

The layer-resolved $B$-O-$B$ bond angles, visualized in Fig.~\ref{fig:BondAngles}, allow to quantify the octahedral rotations.
For bulk LNO strained biaxially to fit the lattice constant of STO,
\textit{ab initio} calculations give in-plane (out-of-plane) bond angles of $165^{\circ}$ ($159^\circ$)
in excellent agreement with the experimental values of $166^{\circ}$ ($160^\circ$)~\cite{May:10}.
In the LNO region, the out-of-plane bond angles $\theta_3$ are always larger than the strained-bulk reference value ($159^\circ$) for all SLs,
whereas the in-plane bond angles $\theta_1$ and $\theta_2$ vary in an interval of $165 \pm 3^{\circ}$.
The deviations from $180^\circ$ are usually stronger in the LNO region than in the STO region.
Moreover, the bond angles are found to depend significantly on the interface polarity
and are more strongly reduced in the $n$-type SL than in the $p$-type SL.
In line with this observation,
the out-of-plane bond angles are larger at IF-$p$ than at IF-$n$ in the $np$-type SL.
A similar enhancement of octahedral tilts within the STO region as a result of electrostatic doping
at $n$-type interfaces was predicted in LAO/STO SLs.~\cite{Doennig:15}

\begin{table}[b]
	\centering
	\caption{\label{tab:Distances-Ni-O}The six out-of-plane Ni-O distances, moving in $[001]$ direction from one interface (IF) to the other ($np$-type SL: from IF-$p$ to IF-$n$). The reference value is $1.94~\AA$ (for both unstrained LNO~\cite{GarciaMunoz:92} and LNO biaxially strained to STO~\cite{May:10}).}
	\begin{ruledtabular}
	\begin{tabular}{lccc}
		  & $d_\text{Ni-O}$ @ first IF & $d_\text{Ni-O}$ @ LNO-C & $d_\text{Ni-O}$ @ second IF	\\
		\hline
		$n$-type SL				& $2.03$ / $1.97$ & $1.92$ / $1.92$ & $1.97$ / $2.03$	\\
		$p$-type SL				& $1.89$ / $1.93$ & $1.98$ / $1.98$ & $1.93$ / $1.89$	\\
		$np$-type SL			& $1.93$ / $1.94$ & $1.95$ / $1.97$ & $1.93$ / $1.97$	\\
	\end{tabular}
	\end{ruledtabular}
\end{table}

A distinct feature of the SLs is the La-Sr distance ($d_{\text{La-Sr}}$ in Fig.~\ref{fig:AtomicStructure}),
which is by $7.7\,\%$ larger in the $n$-type SL ($4.06~\AA$) than in the $p$-type SL ($3.77~\AA$).
Thus, the La-Sr distance can be used as a fingerprint to determine the interface type (electron- or hole-doped),
e.g., in transmission electron microscopy measurements.~\cite{ZhangKeimer:14}
Moreover, there is a considerable buckling in the LaO and SrO interface layers,
where the $A$~cation is found to relax towards the LNO region
by $\Delta z_{\text{$A$-O}}=0.22$ and $0.02~\AA$ for the $n$- and the $p$-type SL
and $0.12$ and $0.09~\AA$ near IF-$n$ and IF-$p$ in the $np$-type SL, respectively.

The in-plane Ni-O bond lengths vary between $1.95$ and $1.97~\AA$ for all SLs, constrained by the lateral lattice constant of STO.
In contrast, the out-of-plane Ni-O bond lengths show much stronger variation, as summarized in Table~\ref{tab:Distances-Ni-O}.
Near an $n$-type (a $p$-type) interface they are by $4.6\,\%$ larger ($2.6\,\%$ smaller) than the bulk value $1.94~\AA$;
hence, the NiO$_6$ octahedra are elongated (compressed) in the $[001]$ direction (note that LNO is exposed to tensile lateral strain).
This effect is less pronounced for the $np$-type SL.
We conclude that the electrostatic doping in the $n$-type ($p$-type) SL
is  responsible
for the increased (decreased) out-of-plane Ni-O bond lengths near the interfaces.
It will become clear in the following that this is associated with
the occupation of Ni-$3d_{z^2}$-derived quantum well (QW) states, which are localized near the interfaces.
Overall, similar structural trends were obtained in (LNO)$_4$/(STO)$_3$ SLs grown on LSAT~\cite{Hwang:13}
or in (LNO)$_1$/(STO)$_1$ SLs.~\cite{KimHan:15}

\section{Electronic structure}

\begin{figure}[tb]
	\centering
	\includegraphics[]{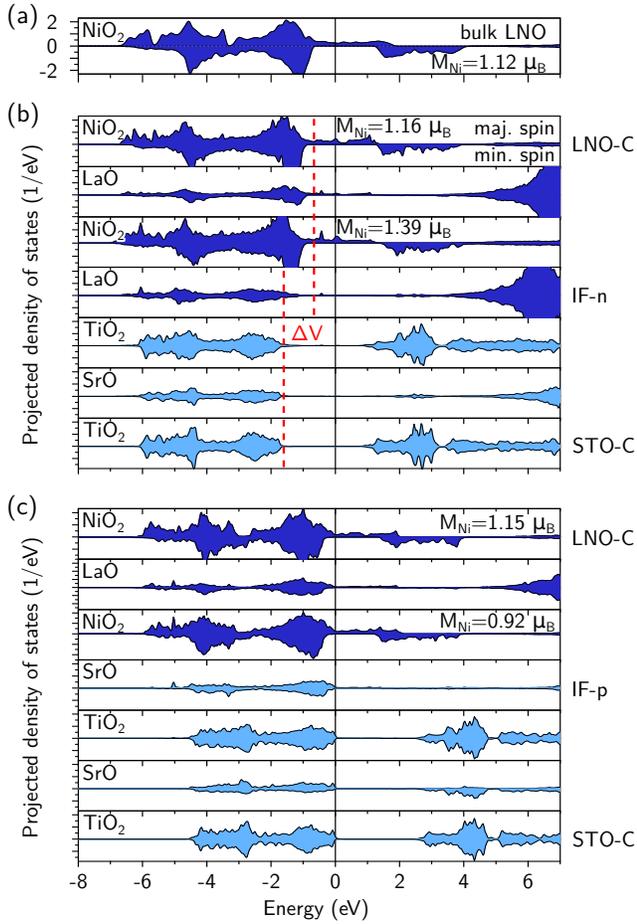}
	\caption{(Color online) Layer- and spin-resolved electronic DOS of the $n$-type~(b) and the $p$-type~(c) SL, together with the DOS of a NiO$_2$ layer in bulk LNO for comparison~(a). Colors and labels refer to Fig.~\ref{fig:AtomicStructure}, and $M_\text{Ni}$ denotes the local magnetic moments at the Ni sites. With respect to the band alignment in the $p$-type SL a potential shift of $\Delta V \approx 1$~eV arises in the $n$-type SL (indicated by the dashed red lines). In addition, the Fermi energy (zero energy) is shifted because of the additional electrons. Note that only half of the layers is shown in each case due to the inversion symmetry of the two SLs.}
	\label{fig:ElectronicStructure-Layers-n-p}
\end{figure}

\begin{figure}[tb]
	\centering
	\includegraphics[]{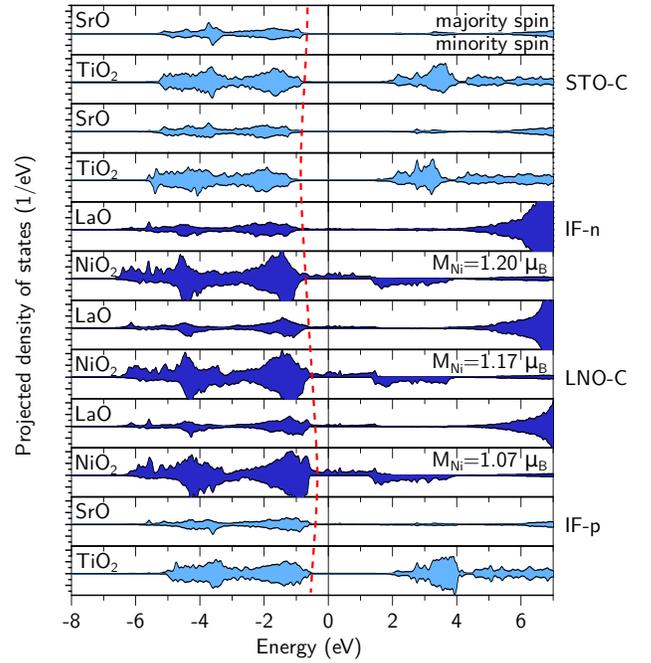}
	\caption{(Color online) Layer- and spin-resolved electronic DOS of the coupled $np$-type SL. Colors and labels refer to Fig.~\ref{fig:AtomicStructure}, and $M_\text{Ni}$ denotes the local magnetic moments at the Ni sites. The dashed red line is a guide to the eye that highlights the modulation of the local potential (amplitude roughly $0.5$~eV) due to the internal electric field.}
	\label{fig:ElectronicStructure-Layers-np}
\end{figure}

The layer- and spin-resolved electronic DOS of the $n$- and the $p$-type SL is shown in Fig.~\ref{fig:ElectronicStructure-Layers-n-p},
together with the DOS of a NiO$_2$ layer in bulk LNO for comparison.
We note that the overall occupation of $e_{g}$ states in bulk LNO is higher than expected from the formal $3+$ oxidation state,
corresponding to a $3d^8 \underline{L}$ configuration rather
than to $3d^7$ ($t_{2g}^6, e_{g}^1$).~\cite{ABR:11, Freeland:11, LiuChakhalian:11, Middey:14}

In the SLs, the electrostatic doping modulates
the occupation of the majority spin Ni~$3d$~$e_{g}$ states,
which are located around the Fermi energy~$E_{\text{F}}$:
Compared with bulk LNO, the $n$-type SL contains additional electrons due to the surplus (LaO)$^{+}$ layer.
In conjunction with the LNO/STO band alignment,
this places $E_{\text{F}}$ within the STO band gap more than $1.7$~eV above the STO valence band and about $1$~eV below the STO conduction band.
This compares well with the valence band offset of $1.75$~eV
reported by Conti \textit{et~al.}\ for (LNO)$_4$/(STO)$_3(001)$ SLs on LSAT.~\cite{Conti:13}
In contrast, for the $p$-type SL $E_{\text{F}}$ coincides with the valence band maximum (VBM) of STO.
Hence, the different interface polarity alters the band alignment between the LNO and the STO region
by $\Delta V \approx 1$~eV (cf.~Fig.~\ref{fig:ElectronicStructure-Layers-n-p}).

Figure~\ref{fig:ElectronicStructure-Layers-np} shows the
DOS of the coupled $np$-type SL.
One can clearly observe the formation of an internal electric field along the $[001]$ direction between the oppositely charged interfaces that leads to shifts of the bands and, consequently, of the local potential with an amplitude of roughly $0.5$~eV.
Built-in electric fields near the LNO interface to Nb-doped STO (Schottky barrier) have been measured recently by means of cross-sectional scanning tunneling spectroscopy.~\cite{Chien:16}

\begin{figure*}[tb]
	\centering
	\includegraphics[]{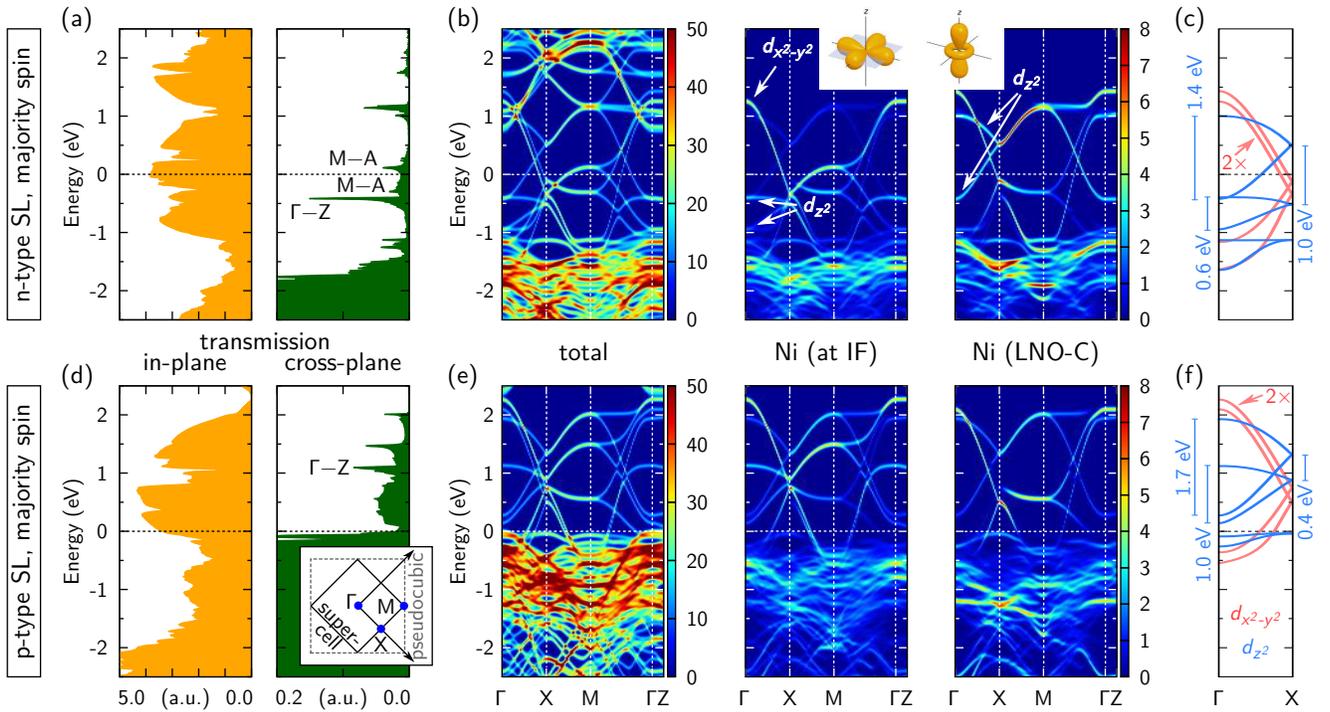}
	\caption{(Color online) Majority-spin electronic transmission ${\cal T}_\uparrow(E)$~(a)+(d) and $\Vek{k}$-resolved total and layer-by-layer projected DOS~(b)+(e) for the $n$-type~(a)+(b) and the $p$-type~(d)+(e) SL. In the $\Vek{k}$-resolved DOS, each electronic state $\varepsilon_{i,\Vek{k},\uparrow}$ is represented by a broadened delta distribution of weight one (total) or a weight equal to the projection of the corresponding wave function onto Ni~$3d$ atomic orbitals (projected). The color scales are in units of $1/$eV. The inset clarifies the relation between the \mbox{(pseudo-)}cubic and our supercell Brillouin zone. Panels~(c) and~(f) emphasize schematically the Ni~$3d$~$e_{g}$ states along the $\Gamma-X$ path.}
	\label{fig:KPDOS-n-p}
\end{figure*}

Such an internal electric field emerges also
in LAO/STO bilayers and SLs.~\cite{Ohtomo:2004,PentchevaPickett:09,LAOSTO-IsmailBeigi:10,PentchevaPickett:06,PentchevaPRL:10}
However, a major difference is that the compensation at the polar interface in LAO/STO systems
involves an occupation of Ti~$3d$ states leading to the formation of a two-dimensional electron gas at the interface,
whereas in LNO/STO SLs STO remains insulating and the electrostatic doping is mainly accommodated in the metallic NiO$_2$ layers. 
This trend is reflected in the Ni magnetic moments, which are always enhanced (reduced) near IF-$n$ (IF-$p$) with respect to the central LNO-C layer.

\section{Electronic transport, Fermi surfaces, and thermoelectric properties}

\subsection{Electronic transport}

\begin{figure*}[tb]
	\centering
	\includegraphics[]{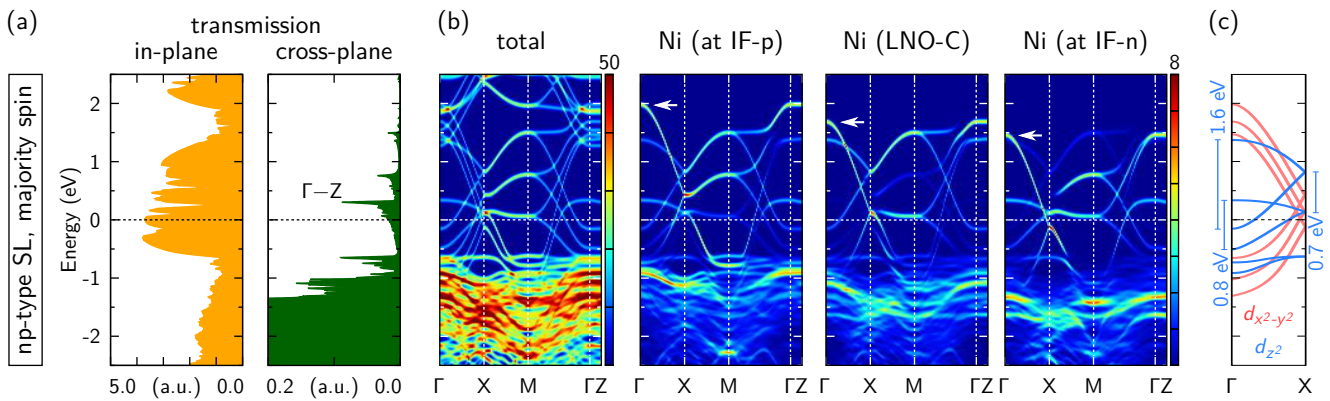}
	\caption{(Color online) Majority-spin electronic transmission ${\cal T}_\uparrow(E)$~(a) and $\Vek{k}$-resolved total and layer-by-layer projected DOS~(b) for the $np$-type SL. The color scales are the same as in Fig.~\ref{fig:KPDOS-n-p}. The small white arrows underline how the Ni~$3d_{x^2-y^2}$ states follow the internal electric field. Panel~(c) emphasizes schematically the Ni~$3d$~$e_{g}$ states along the $\Gamma-X$ path.}
	\label{fig:KPDOS-np}
\end{figure*}

Aiming for thermoelectric properties,
we have calculated the transmission ${\cal T}_\sigma(E)$
in the $[1\bar{1}0]$ and $[110]$ (in-plane) as well as in the $[001]$ (cross-plane) direction for all three SLs.
The two in-plane transmission curves are always equal within the numerical accuracy;
hence, we show only one averaged in-plane curve per SL (cf.~Figs.~\ref{fig:KPDOS-n-p} and~\ref{fig:KPDOS-np}).
Next to the transmission panels we provide the electronic band structure,
which facilitates the understanding of the transmission curves.
In addition, we projected out the Ni~$3d$ contribution for the three different NiO$_2$ layers in the $[001]$ direction.
Since our supercells are rotated in-plane by $45^\circ$
with respect to the conventional unit cell of the \mbox{(pseudo-)}cubic perovskite structure,
the $X$ and $M$~points lie in the $[1\bar{1}0]$ and $[100]$ directions, respectively (see inset in Fig.~\ref{fig:KPDOS-n-p}).
The many band crossings that can be observed in the band structure plots
necessitate a very dense $\Vek{k}$-point grid
to avoid artifacts from the calculation of the band derivatives in the transmission results.
In the following we discuss the properties of the transmission curves around $\mu(T) \approx E_{\text{F}}$ (chosen as zero energy reference here),
which is the relevant region for the thermoelectric quantities.
We restrict our discussion largely to the majority spin channel, since,
with the only exception of the $p$-type SL,
the transport in the minority spin channel is blocked due to band gaps within a $\pm 0.5$~eV window around $E_{\text{F}}$.

The in-plane transmission is metallic for all SLs mainly due to the highly dispersive Ni-$3d$-derived bands,
as one can infer from the projected band structure plots in Figs.~\ref{fig:KPDOS-n-p} and~\ref{fig:KPDOS-np}.
In the case of the $p$-type SL, an additional contribution to the transmission (in both spin channels)
stems from the valence band of STO (below $E_{\text{F}}$);
however, above $E_{\text{F}}$ the band transmission associated with the NiO$_2$ layers clearly dominates.

The cross-plane transmission is purely due to tunneling for the $n$- and the $np$-type SL,
since the Fermi energy lies within the band gap of STO (cf.~Figs.~\ref{fig:ElectronicStructure-Layers-n-p} and~\ref{fig:ElectronicStructure-Layers-np}).
Consequently, it is almost two orders of magnitude smaller than the in-plane transmission.
For the $p$-type SL, in which $E_{\text{F}}$ coincides with the VBM of STO,
charge carriers can pass the STO region via the STO valence band instead of tunneling,
which leads to a strongly increased transmission.
A similar enhancement of the transmission occurs in the minority spin channel, but only below $E_{\text{F}}-0.25$~eV
due to the band gap in the LNO region [cf.~Fig.~\ref{fig:ElectronicStructure-Layers-n-p}(c)].
For all SLs one observes a drastic increase in the majority spin cross-plane transmission as the STO valence band comes into play.
In our approach, cross-plane tunneling transport is accounted for
via the residual dispersion of our \textit{ab initio} band structure along the $[001]$ direction.
Within the tunneling regime, several transmission peaks are observed.
Some arise due to transport along $\Gamma-Z$.
The bands that correspond to these peaks
have a strong Ni~$3d_{z^2}$ character, but no Ni~$3d_{x^2-y^2}$ character.
Other peaks arise due to transport along $M-A$, where $A=(\pi/a, \pi/a, \pi/c)$.
The bands that correspond to these peaks
have a mixed Ni~$3d_{z^2}$ and $3d_{x^2-y^2}$ character.

\subsection{Orbital-selective quantization of the QW states}

Some interesting observations can be made in the band structure plots shown in Figs.~\ref{fig:KPDOS-n-p} and~\ref{fig:KPDOS-np}
that refine the understanding of the electronic structure we have gained
from Figs.~\ref{fig:ElectronicStructure-Layers-n-p} and~\ref{fig:ElectronicStructure-Layers-np}.
Focusing exemplarily on the $\Gamma-X$ path in the Brillouin zone and the energy region within the STO band gap,
two types of bands can be distinguished:
those with a higher in-plane dispersion (derived from Ni~$3d_{x^2-y^2}$ orbitals hybridized with O~$2p_{x,y}$ orbitals)
and those with a lower dispersion (derived from Ni~$3d_{z^2}$ orbitals, which point along the $[001]$ direction).
This is shown schematically in panels~(c) and~(f) in Figs.~\ref{fig:KPDOS-n-p} and~\ref{fig:KPDOS-np}.
Since there are always three NiO$_2$ layers, we find three (folded) bands of each type.

The Ni-$3d_{x^2-y^2}$- and Ni-$3d_{z^2}$-derived states respond differently to the quantum confinement.
The width of the former is always about $2.8$~eV for all three SLs.
Their energy offset follows clearly the local potential, which is particularly obvious for the $np$-type SL and marked by small white arrows in Fig.~\ref{fig:KPDOS-np}(b).
Hence, each band is confined to one single NiO$_2$ layer. Note that the interface bands are energetically degenerate for the symmetric $n$- and $p$-type SL.

The Ni-$3d_{z^2}$-derived bands show a distinctly different behavior,
since the Ni~$3d_{z^2}$ orbitals form QW states across all three NiO$_2$ layers.
The three bands are separated by energies ranging from $0.4$ to $1.0$~eV.
The lowest is predominantly localized in the LNO-C layer,
whereas the second is derived almost exclusively from Ni~$3d_{z^2}$ orbitals at the interfaces
and the topmost shows contributions from all NiO$_2$ layers.
Bandwidth, separation, and occupation of these bands [cf.~Figs.~\ref{fig:KPDOS-n-p}(c),(f) and~\ref{fig:KPDOS-np}(c), blue lines]
strongly depend on the type of interfaces in the SLs.
While for the $p$-type SL only the lowest band is occupied,
for the $np$-type SL the second band is partially occupied and,
finally, for the $n$-type SL the two lowest bands are completely and the topmost band partially filled.
With increasing occupation of the second Ni~$3d_{z^2}$ band, its bandwidth as well as that of the topmost band are reduced.  
Moreover, the higher occupation of the second Ni~$3d_{z^2}$ band correlates
with an enhancement of the out-of-plane Ni-O bond lengths at the interface (cf.~Table~\ref{tab:Distances-Ni-O}).

The analogy to the lowest three states in a finite QW model and their probability densities is obvious.
The energy separation of the three Ni~$3d_{z^2}$ bands is a superposition of quantum confinement
and, in particular, Coulomb repulsion (an increasing occupation of the second band pushes the topmost band to higher energies).
Related effects of orbital-selective quantization have been observed experimentally in metallic SrVO$_3$ thin films on STO.~\cite{Yoshimatsu:11}

\subsection{Fermi surfaces}

\begin{figure}[tb]
	\centering
	\includegraphics[]{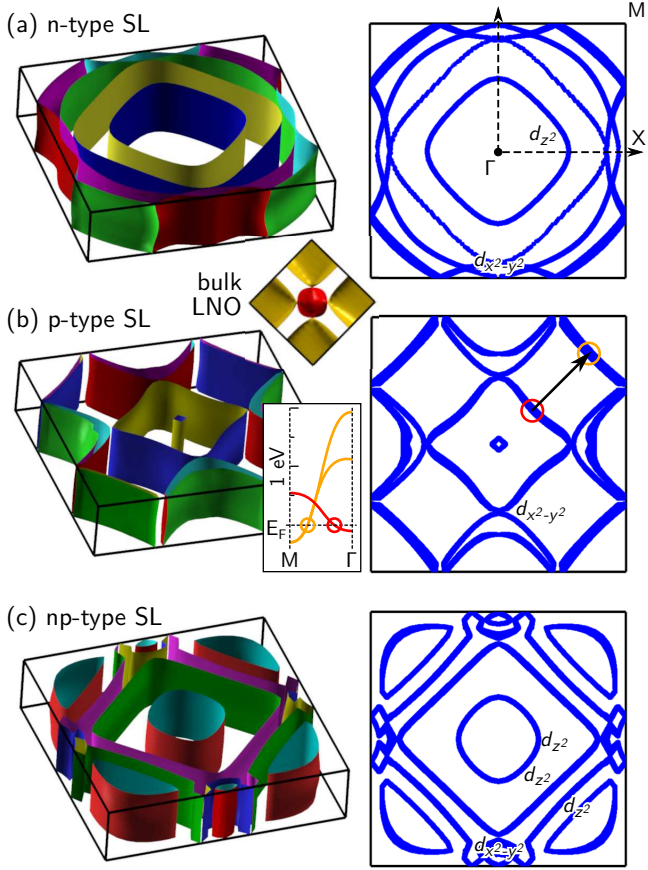}
	\caption{(Color online) Fermi surfaces for the $n$-type~(a), the $p$-type~(b), and the $np$-type SL~(c) in three dimensions (left) and projected onto the $(001)$ plane (right). For some sheets the predominant orbital character is noted at the corresponding positions in the Brillouin zone. Fermi surface nesting is indicated for the $p$-type SL together with a schematic fraction of the band structure [cf.~Fig.~\ref{fig:KPDOS-n-p}(e)]. The Fermi surface of bulk LNO is shown for comparison.}
	\label{fig:FermiSurfaces}
\end{figure}

\begin{figure*}[tb]
	\centering
	\includegraphics[]{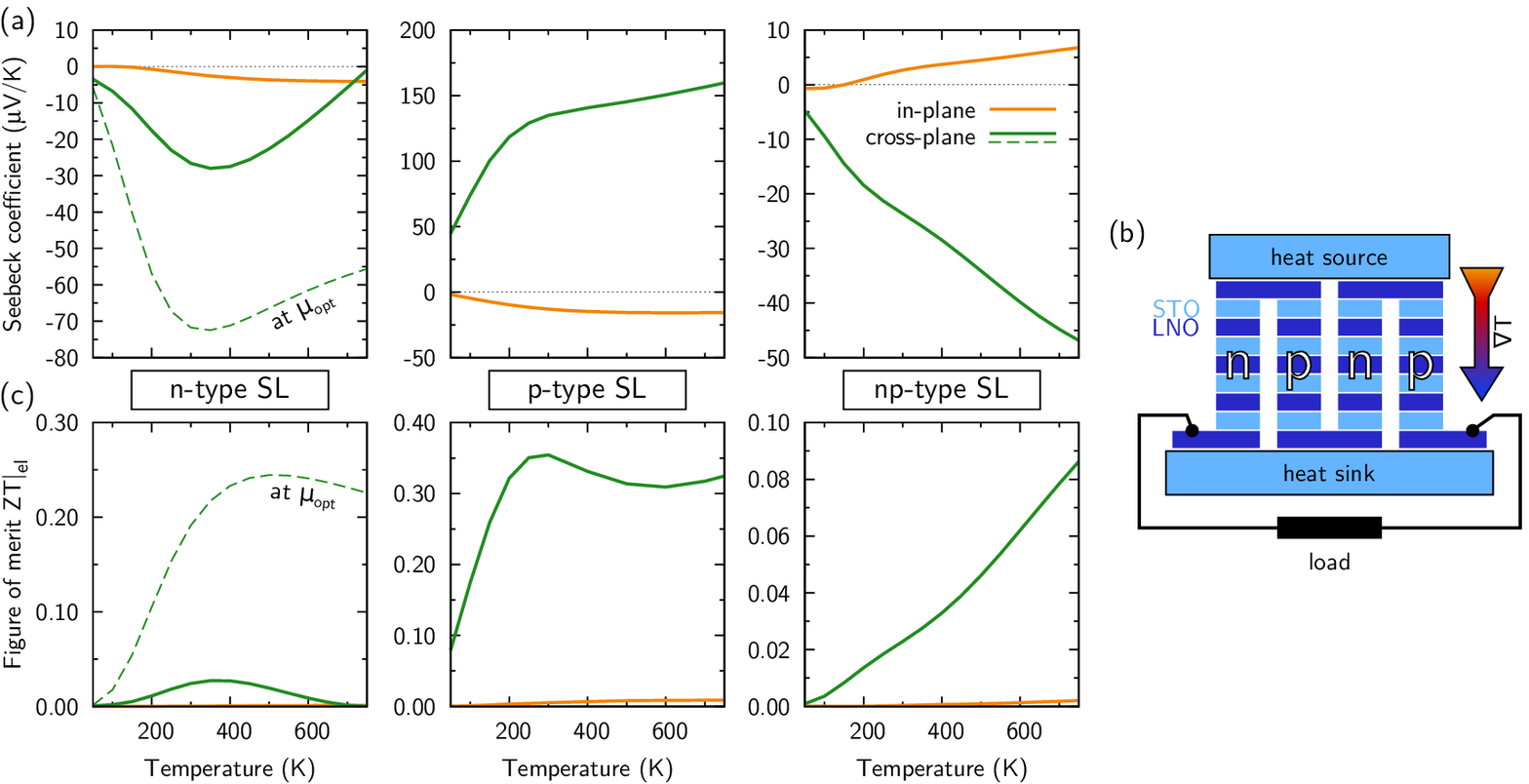}
	\caption{(Color online) (a)~Effective Seebeck coefficients $S_{\text{eff}}(T)$ for the three different SLs. In-plane and cross-plane results are depicted by solid orange and green lines, respectively. (b)~Illustration of an oxide-based thermoelectric generator constructed from $n$- and $p$-type SLs. (c)~Figure of merit $ZT \vert_{\text{el}}$ for the three different SLs. Thin dashed lines illustrate further improvements attained by doping for the $n$-type SL.}
	\label{fig:Seebeck}
\end{figure*}

The interface polarity has a strong influence on the Fermi surfaces shown in Fig.~\ref{fig:FermiSurfaces}.
A common feature is the quasi two-dimensional cylindrical shape,
in contrast to the three-dimensional shape of bulk LNO (see inset in Fig.~\ref{fig:FermiSurfaces}).
The $n$- and the $np$-type SL have electron pockets around the $\Gamma$~point, formed by the Ni~$3d_{z^2}$ QW states discussed above,
and Ni~$3d_{x^2-y^2}$ features around the $X$~point.
The $\Gamma$-centered electron pocket of the $p$-type SL has a similar diameter,
but strong Ni~$3d_{x^2-y^2}$ character along $\Gamma-X$ and mixed character along $\Gamma-M$,
and the pocket around the $M$~point has mixed character as well.
Taking into account the band folding due to the $c(2\times2)$ lateral unit cell, some features are comparable
to calculated Fermi surfaces for a (LNO)$_1$/(LAO)$_1(001)$ SL~\cite{Hansmann:09, Puggioni:12}
or to photoemission results on LNO epitaxial films.~\cite{LNO-STO-FermiSurfaces:09, King:14, LNO-Instab:2015, Yoo:16}
Overall, the more complex structure and topology of our LNO/STO SL Fermi surfaces is caused
by the potential- and confinement-induced band splitting discussed above.

For epitaxial LNO thin films under tensile strain, experiments suggest the possibility of enhanced Fermi surface nesting.~\cite{LNO-Instab:2015}
Figure~\ref{fig:FermiSurfaces} shows that not only strain, but also the interface polarity plays an important role
in controlling the degree of possible Fermi surface nesting (note the parallel segments of the squarelike inner and outer sheet of the $p$-type SL).

\subsection{Thermoelectric properties}

Finally, we discuss the thermoelectric properties.
Since the temperature dependence of the chemical potential can play an important role,~\cite{Geisler-Heusler:15}
we have calculated $\mu(T)$ from the (rigid) electronic structure (see Appendix)
and included it in the calculation of the Seebeck coefficients, Eq.~\eqref{eq:EffectiveSeebeck},
which are shown in Fig.~\ref{fig:Seebeck}(a).
For all three SLs, the cross-plane Seebeck coefficient is much larger than the in-plane Seebeck coefficient (in absolute value).
The $p$-type SL exhibits a particularly large positive cross-plane Seebeck coefficient
of $S_{\text{eff}} \approx 135~\mu$V/K at room temperature.

Sign and magnitude of the Seebeck coefficient correlate with the slope of the transmission around $\mu(T)$ due to the numerator in Eq.~\eqref{eq:Seebeck}.
Three examples shall illustrate this:
(i)~For the $p$-type SL, the rapid increase of the cross-plane transmission directly below $E_{\text{F}}$
due to the involvement of the STO valence band [cf.~Fig.~\ref{fig:KPDOS-n-p}(d)]
leads to a strong asymmetry (negative slope), which causes the large positive Seebeck coefficient.
(ii)~From the majority spin in-plane transmission for the $p$-type SL one would expect a large negative Seebeck coefficient
due to the positive slope near $E_{\text{F}}$.
However, the transmission in the minority spin channel (not shown) due to the STO valence band has a negative slope,
which leads to a counteracting contribution.
In total, Eq.~\eqref{eq:EffectiveSeebeck} gives a small negative in-plane Seebeck coefficient.
(iii)~For the $n$-type SL, the cross-plane $M-A$ tunneling peak right above the Fermi energy [cf.~Fig.~\ref{fig:KPDOS-n-p}(a)]
leads to a slight positive slope and thus to a negative Seebeck coefficient.

While the cross-plane Seebeck coefficient of the $n$-type SL is close to the measured value for bulk LNO
($S_{\text{eff}} \approx -20~\mu$V/K at room temperature~\cite{LNO-LCO-Thermo:14}),
the large positive cross-plane Seebeck coefficient found for the $p$-type SL shows
that the presented nanostructuring indeed improves the thermoelectric properties of the active material LNO.
More importantly, the possibility to design the thermoelectric response of a SL from $n$-type ($S_{\text{eff}} < 0$) to $p$-type ($S_{\text{eff}} > 0$)
by varying solely the interface polarity [cf.~Fig.~\ref{fig:Seebeck}(a)]
allows for constructing an oxide-based thermoelectric generator, as illustrated in Fig.~\ref{fig:Seebeck}(b).
This strategy avoids typical compatibility issues between different $n$- and $p$-type materials.

Figure~\ref{fig:Seebeck}(c) shows upper bounds for the dimensionless thermoelectric figure of merit,
in which the phonon contribution to the thermal conductivity $\kappa_{\text{ph}}$ has been neglected
and the electronic contribution $\kappa_{\text{el}}$ has been calculated by using Eq.~\eqref{eq:ThermalConductivityEl}:
\begin{equation*}
	ZT \vert_{\text{el}} = S_{\text{eff}}^2 \, \sigma \, T / \kappa_{\text{el}}
	\text{.}
\end{equation*}
The $p$-type SL can reach a cross-plane value of $ZT = 0.35$ around room temperature, which is considerable for oxide materials.
The values for the $n$- and the $np$-type SL are one order of magnitude smaller,
and the in-plane values are generally negligible, but in line with
$ZT \vert_{\text{el}} = 0.016$ for bulk LNO at room temperature
estimated from experimental data.~\cite{LNO-LCO-Thermo:14}
Hence, the thermoelectric performance is significantly increased for the $p$-type SL,
but rather poor for the $n$-type SL.
One route to improve $S_{\text{eff}}$ and $ZT$ of the $n$-type SL is by additional doping:
A rigid shift of the chemical potential to $\mu_{\text{opt}} = E_{\text{F}} - 0.5$~eV
[just below the $\Gamma-Z$ tunneling peak, cf.~Fig.~\ref{fig:KPDOS-n-p}(a)]
leads to a three times higher cross-plane Seebeck coefficient and a figure of merit comparable to the one of the $p$-type SL (cf.~Fig.~\ref{fig:Seebeck}).

A more precise estimate for the figure of merit requires
calculation of $\kappa_{\text{ph}}$ \textit{ab initio}, which is beyond the scope of this work.
We only note the following aspects here:
(i)~$\kappa_{\text{ph}}$ will be anisotropic due to the interface thermal resistance:
Since La is much heavier than Sr, phonon mismatch and scattering are likely to cause a beneficial reduction in particular
of the cross-plane $\kappa_{\text{ph}}$.~\cite{Holuj:15}
(ii)~In addition, the relaxation time $\tau$ is required.
It is typically in the femtosecond range~\cite{Garcia:12}
and can differ by orders of magnitude between in- and cross-plane transport.~\cite{Gruner:15}

\section{Summary}

By using DFT+$U$ calculations,
we investigated the influence of the interface polarity
on the lattice and electronic structure
as well as the electronic transport and thermoelectric properties
of epitaxial LaNiO$_3$/SrTiO$_3(001)$ superlattices on a SrTiO$_3$ substrate.
The band alignment was found to depend strongly on the type of interfaces in the SL,
i.e., $n$-type (LaO)$^{+}$/(TiO$_2$)$^{0}$ or $p$-type (NiO$_2$)$^{-}$/(SrO)$^{0}$.
In contrast to the LaAlO$_3$/SrTiO$_3$ system,
the $n$- and $p$-type doping is solely accommodated by changing the occupation of bands within the LaNiO$_3$ region
and determines the position of the Fermi energy within the SrTiO$_3$ band gap.
Coupling interfaces of opposite polarity generates an electric field in the superlattice.
We explained the electronic structure within the SrTiO$_3$ band gap
in terms of an orbital-selective quantization of quantum well states in the metallic LaNiO$_3$ region
(Ni~$3d_{x^2-y^2}$ vs.\ $3d_{z^2}$)
together with Coulomb repulsion effects.
The potential- and confinement-induced splitting of the bands leads to complex Fermi surfaces of the differently doped superlattices
with a quasi two-dimensional cylindrical shape and a tendency towards Fermi surface nesting that depends on the interface polarity.
The octahedral rotations of LaNiO$_3$ were found to carry over to the SrTiO$_3$ layers.
The size of octahedral tilts and La-Sr spacings across the interface depends on the interface polarity.
Likewise, the Ni-O out-of-plane bond lengths and the Ni magnetic moments at the interfaces increase with the electron doping
due to successive occupation of Ni-$3d_{z^2}$-derived quantum well states.

Analysis of electronic transport and thermoelectric properties based on Boltzmann theory in the constant relaxation time approximation
showed anisotropic Seebeck coefficients and figures of merit.
A particularly large Seebeck coefficient
($135~\mu$V/K at room temperature)
and the highest figure of merit ($0.35$)
were found for cross-plane transport through the $p$-type superlattice,
which was attributed to the involvement of the SrTiO$_3$ valence band.
Finally, we showed that a selective design of $n$- and $p$-type oxide thermoelectrics is possible,
exploiting the interface polarity in an oxide superlattice.
This provides a route for constructing an oxide-based thermoelectric generator,
avoiding possible compatibility issues between different $n$- and $p$-type materials.
While the LaNiO$_3$/SrTiO$_3$ SLs with their moderate size of Seebeck coefficient and figure of merit serve mainly as a proof of principle here,
further strategies like $n$-type doping of the SrTiO$_3$ part as well as
different materials combinations with improved characteristics need to be explored in future studies.

\begin{acknowledgments}

We thank B.~Keimer and H.-U.~Habermeier (Stuttgart) as well as W.~E.~Pickett (Davis) for discussions
and M.~E.~Gruner (Duisburg) for helpful comments on our manuscript.
This work was supported by the German Science Foundation (Deutsche Forschungsgemeinschaft, DFG) within the SFB/TRR~80, projects G3 and G8.

\end{acknowledgments}

\appendix*

\section{Obtaining electronic transport and thermoelectric properties from the DFT electronic structure}

In the regime of linear response,
in which temperature gradients and voltages that are applied to the SLs are assumed to be small,
the thermoelectric properties can be obtained by using the approach of Sivan and Imry.~\cite{SI86}
The central quantity in this approach is the energy- and spin-resolved transmission ${\cal T}_\sigma(E)$
(also known as transport distribution~\cite{MahanSofo:96}),
which we calculate by using the the BoltzTraP code.~\cite{BoltzTraP:06}

From the \textit{ab initio} electronic structure $\varepsilon_{i,\Vek{k},\sigma}$
we start by calculating the group velocities in different directions $\Vek{e}$,
\begin{equation*}
	v_{i,\Vek{k},\sigma} = \frac{1}{\hbar} \ \Vek{e} \Multp \Vek{\nabla}_k \ \varepsilon_{i,\Vek{k},\sigma}
	\text{,}
\end{equation*}
which we use to define the energy- and spin-resolved transmission in the corresponding direction,
\begin{equation*}
	{\cal T}_\sigma(E) = \frac{e^2}{N} \sum_{i,\Vek{k}} \delta(E-\varepsilon_{i,\Vek{k},\sigma}) \left( v_{i,\Vek{k},\sigma} \right)^2
	\text{,}
\end{equation*}
where $N$ is the total number of calculated $\Vek{k}$~points.
Within the common approximation of constant relaxation time $\tau$, the electrical conductivity can be expressed as
\begin{equation}
	\label{eq:Conductivity}
	\sigma_\sigma(T)  = -\frac{\tau}{\Omega} \,
		\int \text{d}E \, \frac{\partial f}{\partial E} \,
		{\cal T}_\sigma(E)
	\text{,}
\end{equation}
where $\Omega = a^2 c$ is the volume of the considered supercell
and $f = f_{\mu, T}(E)$ denotes the Fermi distribution function.
The total conductivity is simply $\sigma = \sigma_{\uparrow} + \sigma_{\downarrow}$.
The spin-projected Seebeck coefficients take on the form 
\begin{equation}
	\label{eq:Seebeck}
	S_\sigma(T)  = -\frac{1}{eT} \,
		\frac{{\displaystyle \int \text{d}E \,
		\frac{\partial f}{\partial E} \,
		(E-\mu) \, {\cal T}_\sigma(E)}}
		{{\displaystyle\int \text{d}E \,
		\frac{\partial f}{\partial E} \, {\cal T}_\sigma(E)}}
	\text{.}
\end{equation}
They are \textit{not} additive ($S \neq S_{\uparrow} + S_{\downarrow}$) due to the different denominators
and do not have a strict physical meaning.
However, with these quantities
the effective (charge) Seebeck coefficient can be expressed as
\begin{equation}
  \label{eq:EffectiveSeebeck}
  S_{\text{eff}} = \frac{\sigma_\uparrow\,S_\uparrow +
                         \sigma_\downarrow\,S_\downarrow}
                    {\sigma_\uparrow + \sigma_\downarrow}
\text{,}
\end{equation}
treating the two spin channels as parallel connected resistors.
In contrast to the conductivity, the effective Seebeck coefficient does \textit{not} depend on the value of the relaxation time.

It is helpful to note
that the Seebeck coefficient defined in Eq.~\eqref{eq:Seebeck}
measures the asymmetry of the transmission ${\cal T}_\sigma(E)$ around the chemical potential $\mu$
due to its numerator.
For instance, it vanishes for zero slope and takes on positive (negative) values for negative (positive) slope of ${\cal T}_\sigma(E)$.

\begin{figure}[tb]
	\centering
	\includegraphics[]{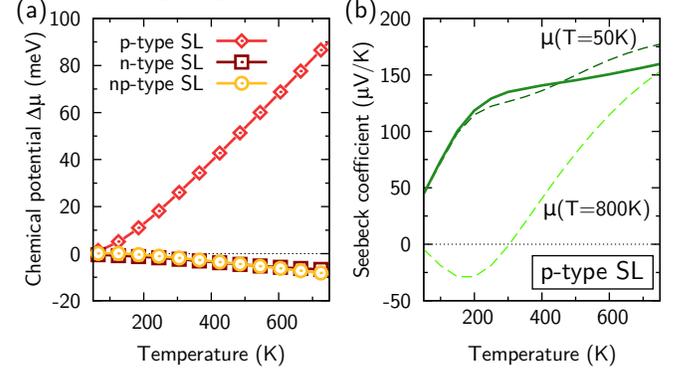}
	\caption{(Color online) (a)~Calculated variation of the chemical potential $\mu(T) - \mu(0)$. (b)~Effective cross-plane Seebeck coefficient $S_{\text{eff}}(T)$ for the $p$-type superlattice (solid line), together with two curves for which $\mu$ has been fixed to its values at $T=50$~K and $T=800$~K (dark and bright thin dashed lines), emphasizing the relevance of the temperature-dependent chemical potential.}
	\label{fig:AppFig}
\end{figure}

The temperature dependence of the chemical potential can play an important role.~\cite{Geisler-Heusler:15}
We have therefore calculated $\mu(T)$ from the (rigid) electronic structure [cf.~Fig.~\ref{fig:AppFig}(a)].
For the $n$- and the $np$-type superlattice, the variation is below $10$~meV in the temperature range considered here.
However, the proximity of the SrTiO$_3$ valence band leads to a ten times larger variation for the $p$-type superlattice.
In addition to the cross-plane Seebeck coefficient that includes $\mu(T)$ [cf.~Fig.~\ref{fig:Seebeck}(a)],
Fig.~\ref{fig:AppFig}(b) contains two curves for fixed $\mu$ at $T=50$~K and $T=800$~K.
For these two curves, the temperature dependence of the Seebeck coefficient solely stems
from the broadening of the Fermi distribution function in Eqs.~\eqref{eq:Conductivity} and~\eqref{eq:Seebeck}.
The deviations underline the importance of accounting for the temperature dependence of the chemical potential.

Finally, the transmission provides also the electronic contribution to the thermal conductivity,
\begin{equation}
	\label{eq:ThermalConductivityEl}
	\kappa_{\text{el}}(T)  = -\frac{\tau}{e^2 T \Omega} \,
		\int \text{d}E \, \frac{\partial f}{\partial E} \, (E-\mu)^2 \,
		\left\lbrace {\cal T}_\uparrow(E) + {\cal T}_\downarrow(E) \right\rbrace
	\text{,}
\end{equation}
which enters the thermoelectric figure of merit~$ZT$.

Alternatives to Boltzmann theory used here
include the semiclassical Wentzel-Kramers-Brillouin approximation~\cite{Ricci:13}
or pure quantum transport~\cite{Geisler-Heusler:15, GeislerPopescu:14, ComtesseGeisler:14, Popescu:13}
based on an \textit{ab initio} potential.


\end{document}